\documentstyle[11pt,AATS,epsf]{article}	

\begin{document}	

\title{Observing the planet formation time-scale by ground-based 
direct imaging of planetary companions to young nearby 
stars: Gemini/$H\bar ok\bar upa'a$ image of TWA-5} 

\author{Ralph Neuh\"auser}
\affil{IfA, University of Hawai'i, USA \& MPE, D-85740 Garching, Germany}
\author{Dan Potter}
\affil{Institute for Astronomy, University of Hawai'i, 2680 Woodlawn Dr., Honolulu HI 98622, USA}
\author{Wolfgang Brandner}
\affil{IfA, University of Hawai'i, USA \& ESO, D-85748 Garching, Germany}


\begin{abstract}
Many extra-solar planets and a few planetary systems have been found 
indirectly by small periodic radial velocity variations around 
old nearby stars. The orbital characteristics of
most of them are different from the planets in our solar system.
Hence, planet formation theories have to be revised.
Therefore, observational constraints regarding young planets would
be very valuable. We have started a ground-based direct imaging search
for giant planets in orbit around young nearby stars.
Here, we will motivate the sample selection
and will present our direct imaging observation
of the very low-mass (15 to 40 Jupiter masses) brown dwarf companion
TWA-5 B in orbit around the nearby young star TWA-5 A, recently
obtained with the 36-element
curvature-sensing AO instrument $H\bar ok\bar upa'a$ 
of the University of Hawai'i
at the 8.3m Gemini-North telescope on Mauna Kea.
We could achieve a FWHM of 64 mas and $25~\%$ Strehl.
We find significance evidence for orbital motion of B around A.
\end{abstract}


\section{Introduction: Direct imaging search for giant planets}

So far, no direct imaging detection of an extra-solar planet in orbit around a
star has been achieved, mainly because of the problem of dynamic range:
Planets are too faint and too close to their bright primary stars.
Neither speckle techniques nor space-based observations have been able
to directly detect a planet around another star.
From radial velocity observations, it is known, though, that planets
and even planetary systems do exist around other stars.
One can avoid the problem of dynamic range by observing young nearby
stars, where there could be young planets still contracting and 
accreting, so that they are relatively hot and (self-)luminous,
e.g. Burrows et al. (1997).

A well-suited sample for such a program is the TW Hya association (TWA) of
a few dozen young ($\sim 10$ Myr) low-mass pre-main sequence (i.e. T Tauri)
stars at a distance of roughly 55 pc (e.g. Webb et al. 1999).
Several members of TWA have been observed by the HST
NICMOS, where two planet candidates and one brown dwarf companion
candidate were detected:
A planet candidate near TWA-7 (9.5 mag fainter than the star in H and K
at a separation of 2.5 arc sec) was also detected by H- and K-band speckle 
from the ground (Neuh\"auser et al. 2000a); an H-band spectrum taken with 
ISAAC at the VLT has
shown that it is a background K-type star (Neuh\"auser et al. 2001).
The planet candidate near TWA-6 has not yet been confirmed nor rejected,
it is 12 mag fainter than the star in H at a separation of 2.5 arc sec
(Schneider et al. 2001).
The brown dwarf companion candidate 2 arc sec north of TWA-5
(presented first by Lowrance et al. 1999 and Webb et al. 1999, 
and also observed by Weintraub et al. 2000), has been confirmed
by both proper motion and spectroscopy (spectral type M9) by
ground-based optical and IR follow-up imaging and spectroscopy 
using FORS2 and ISAAC at the VLT (Neuh\"auser et al. 2000b);
simultaneously and independently, also Schneider et al. (2001)
took a spectrum of this object and confirmed the spectral type
first published by Neuh\"auser et al. (2000b) using an HST
STIS spectrum with smaller wavelength range.

The brown dwarf companion TWA-5 B has a mass of 15 to 40 Jupiter
masses (according to different theoretical tracks and isochrones),
an age of $12 \pm 6$ Myrs (as TWA-5 A), and it is the 4th brown dwarf
companion in orbit around a normal star confirmed by both spectrum and
proper motion. It is the one with the lowest mass, possibly only 
slightly above the deuterium burning mass limit. It is also the first one
in orbit around a pre-main sequence star and the first one in orbit
around a spectroscopic binary: TWA-5 A is a single- or possibly
double-lined SB T Tauri star (Torres et al. 2001).

\section{Gemini $H\bar ok\bar upa'a$ observation of TWA-5}

We observed several TWA members with the University of Hawai'i (UH)
36-element curvature-sensing Adaptive Optics (AO) 
instrument $H\bar ok\bar upa'a$ at the 8.3m Gemini-North 
on Mauna Kea, Hawai'i, in UH pay-back time in February 2001.
TWA-5 was observed in the photometric night 23/24 Feb 2001 at sub-arc 
sec seing condition. We observed the object with the Wollaston,
the dual-imaging polarimeter, in order to detect a possible
circumstellar disk around the star (Potter et al., in prep.).
As usual with $H\bar ok\bar upa'a$, we used the UH IR camera QUIRC.
Individual exposure times were 6 seconds,
obtained at different positions and rotations on the chip.
After dark and sky subtraction and flat fielding, we shifted 
and co-added all frames to the final image with 8 min total 
exposure (fig. 1).  
We could achieve a FWHM of 64 mas and $25~\%$ Strehl.

\begin{figure}[htb]	
\vspace{5cm}
\includegraphics{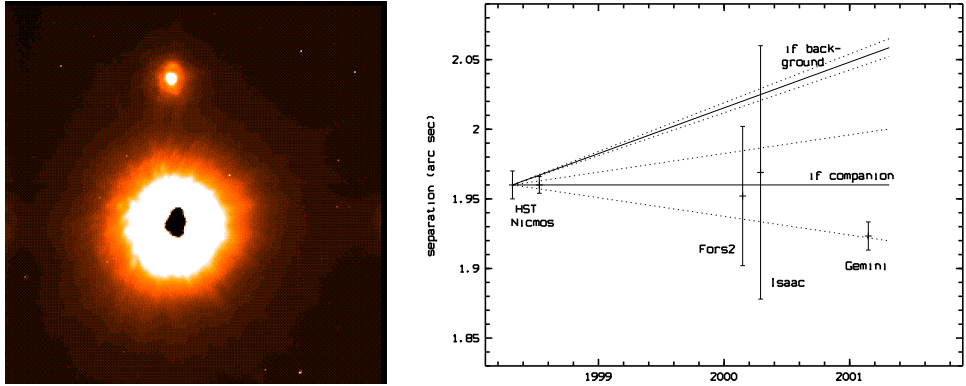}
\caption{left: $H\bar ok\bar upa'a$/QUIRC H-band image of the faint brown
dwarf TWA-5 B (64 mas FWHM) two arc sec north of the bright 
TWA-5 A (with saturated center elongated by 40 mas).
right: Change in separation between A and B from 1998 to 2001,
not consistent with B being a non-moving background object
(proper motion of A is known from Tycho), but with B being a
companion (dotted lines allow for orbital motion,
see Neuh\"auser et al. 2000b).}
\end{figure}

From the similar brightness of the two SB components of TWA-5 A
in high-resolution spectra, both stars of the SB are of simliar 
spectral type, hence both are early M-type stars, like the
type in (spatially unresolved) spectra of TWA-5 A.
The SB orbit has not been solved, yet.
If the elongation seen in the saturated part of TWA-5 A in figure 1
is due to these two stars, then their separation of roughly 50
to 100 mas would correspond to a $\sim 10$ year orbit at 55 pc,
consistent with the radial velocity variation seen since a few years.
It may well be possible within a few years to determine
the masses of both components dynamically, namely by simultaneous
solutions for the astrometric and spectroscopic orbit.

%
%


\acknowledgments We are gratefull to the University of Hawai'i Adaptive Optics
Group for their support and help with the $H\bar ok\bar upa'a$ observations.
RN wishes to thank the University of Hawai'i Institute for Astronomy for 
hospitality during his visit from Oct 2000 to March 2001.


\begin{references}
\reference Burrows A., Marley M., Hubbard W., et al. 1997, \apj, 491, 856
\reference Lowrance P.J., McCarthy C., Becklin E.E., et al., 1999, \apj, 512, L69
\reference Neuh\"auser R., Brandner W., Eckart A., Guenther E.W.,
Alves J., Ott T., Hu\'elamo N., Fern\'andez M., 2000a, A\&A, 354, L9
\reference Neuh\"auser R., Guenther E.W., Petr M.G., Brandner W.,
Hu\'elamo N., Alves J., 2000b, A\&A, 360, L39
\reference Neuh\"auser R., Guenther E.W., Brandner W., Hu\'elamo N., Ott T.,
Alves J., Comer\'on F., Eckart A., Cuby J.-G., 2001,
In: From Darkness to Light, ed. by Montmerle T. \& Andre P.,
ASP Conf. Series, in press, astro-ph/0007305
\reference Schneider G., Becklin E.E., Lowrance P.J., Smith B.A., 2001,
In: Disks, Planets, and Planetesimals,
ASP Conf. Ser., in press, astro-ph/0007330
\reference Torres G., Neuh\"auser R., Latham D.W., 2001, 
In: Young stars near Earth: Progress and Prospects,
ed. by R. Jayawardhana \& T. Greene,
ASP Conf. Series, in press
\reference Webb R.A., Zuckerman B., Platais I., et al. 1999, \apj, 512, L63
\reference Weintraub D.A., Saumon D., Kastner J.H., Forveille T., 2000, \apj, 530, 867
\end{references}
\end{document}